\documentclass{article}

\usepackage[tbtags]{amsmath}
\usepackage{amsfonts,amssymb}

\makeatletter
\@ifundefined{theorem@style}{\RequirePackage{theorem}}{}

\gdef\th@plain{\normalfont\slshape
  \def\@begintheorem##1##2{%
\item[\hskip\parindent\hskip\labelsep\theorem@headerfont ##1\ ##2\unskip.]}%
\def\@opargbegintheorem##1##2##3{%
\item[\hskip\parindent
\ifx\empty##1\else\hskip\labelsep\fi\theorem@headerfont ##1\ ##2\unskip]{\theorem@headerfont{\rm ##3}.} }}

\gdef\th@definition{\normalfont
  \def\@begintheorem##1##2{%
\item[\hskip\parindent\hskip\labelsep\theorem@headerfont ##1\ ##2\unskip.]}%
\def\@opargbegintheorem##1##2##3{%
\item[\hskip\parindent
\ifx\empty##1\else\hskip\labelsep\fi\theorem@headerfont ##1\ ##2\unskip]{\theorem@headerfont{\rm ##3}.} }}

\makeatother
\usepackage{graphicx}

\let\ln=\log
\numberwithin{equation}{section}
\theoremstyle{plain}
\newtheorem{theorem}{Theorem}
\newtheorem{proposition}{Proposition}
\newtheorem{lemma}{Lemma}
\theoremstyle{definition}
\newtheorem{proof}{Proof}

\newtheorem{demo}{Proof of the lemmas}

\newtheorem{examples}{Examples}

\newtheorem{remark}{Remark}

\newcommand{\sign}{\operatorname{sgn}}
\newcommand\ptl{\partial}
\newcommand\vep{\varepsilon}
\newcommand\vfi{\varphi}
\newcommand\R{\mathbb{R}}

\begin{document}


\title{Nonlinear equations for $p$-adic open, closed, and open-closed strings}

\author{V.~S.~Vladimirov\footnotemark[1]}

\footnotetext[1]{Steklov Mathematical Institute, RAS, Moscow, Russia,
e-mail: vladim@mi.ras.ru.}

\date{}

\maketitle

\begin{abstract}
We investigate the structure of solutions of boundary value problems for a
one-dimensional nonlinear system of pseudodifferential equations describing
the dynamics {\rm(}rolling{\rm)} of $p$-adic open, closed, and open-closed
strings for a scalar tachyon field using the method of successive
approximations. For an open-closed string, we prove that the method converges
for odd values of $p$ of the form $p=4n+1$ under the condition that the
solution for the closed string is known. For $p=2$, we discuss the questions
of the existence and the nonexistence of solutions of boundary value problems
and indicate the possibility of discontinuous solutions appearing.
\end{abstract}

\medskip

\hfill{\sl To Anatolii Alekseevich Logunov on his 80th birthday}

\section{Introduction}
\label{sec1}

For describing the tachyon dynamics of open and closed $p$-adic strings, the
Lagrangian
\begin{align}
\mathfrak L={}&\frac1{h^2}\biggl[-\frac{p^4}{2(p^2-1)}\Psi
p^{-\square/4}\Psi+\frac{p^4}{p^4-1}\Psi^{p^2+1}\biggr]+{}
\nonumber
\\[2mm]
&+\frac1{g^2}\biggl[-\frac{p^2}{2(p-1)}\Phi p^{-\square/2}\Phi+
\frac{p^2}{p^2-1}\Psi^{p(p-1)/2}(\Phi^{p+1}-1)\biggr]
\label{1.1}
\end{align}
was suggested~\cite{1}, where $\Psi(t,x)$ and $\Phi(t,x)$,
$x=(x_1,x_2,\dots,x_{d-1})$, are tachyon fields for open and closed strings,
$h$ and $g$ are interaction constants between open and closed string sectors,
$\square=-\ptl^2_t+\nabla^2_x$ is the $d$-dimensional d'Alembertian, and
$p$ is a prime, $p=2,3,5,\dots$. (In what follows, $p$ is assumed to be an
integer greater than one). The corresponding equations of motion following
from~\eqref{1.1} for $\lambda^2=h^2/g^2\ne0$ have the form~\cite{2}
\begin{subequations}
\begin{align}
&\Psi^{p^2}-p^{-\square/4}\Psi+\lambda^2\frac{p-1}{2p}\Psi^{p(p-1)/2-1}
(\Phi^{p+1}-1)=0,
\label{1.2a}
\\[2mm]
&\Phi^p\Psi^{p(p-1)/2}-p^{-\square/2}\Phi=0.
\label{1.2b}
\end{align}
\label{1.2}
\end{subequations}

Passing to the limit as $\lambda^2\to0$ in system of equations~\eqref{1.2},
we obtain the {\sl simplified system of equations for an open-closed
string}~\cite{2}
\begin{subequations}
\begin{align}
&\Psi^{p^2}=p^{-\square/4}\Psi,
\label{1.3a}
\\[2mm]
&\Phi^p\Psi^{p(p-1)/2}=p^{-\square/2}\Phi.
\label{1.3b}
\end{align}
\label{1.3}
\end{subequations}

\noindent
System of equations~\eqref{1.3} includes Eq.~\eqref{1.3a} describing the
dynamics for the field $\Psi$ of a closed string and Eq.~\eqref{1.3b}
describing the dynamics for the field $\Phi$ of an open string for a known
field $\Psi$. For $\Psi=1$, system~\eqref{1.3} becomes the equation for an
open string,
\begin{equation}
\Phi^p=p^{-\square/2}\Phi.
\label{1.4}
\end{equation}

System~\eqref{1.3} has the vacuum solutions
\begin{equation}
\begin{alignedat}{2}
&(\Psi=0,\ \Phi=0)\quad\forall p,&\qquad
&(\Psi=1,\ \Phi=1),\quad p\text{ is even},
\\[2mm]
&(\Psi=1,\ \Phi=\pm1),\quad p=4n+3,&\qquad
&(\Psi=\pm1,\ \Phi=\pm1),\quad p=4n+1.
\end{alignedat}
\label{1.5}
\end{equation}
For $d=1$, system~\eqref{1.3}, Eqs.~\eqref{1.3a}, and~\eqref{1.4}
respectively describe the motion (rolling) of tachyons in time for
open-closed, closed, and open strings. In this case, system~\eqref{1.3}
becomes
\begin{subequations}
\begin{align}
&\Psi^{p^2}=p^{\ptl^2_t/4}\Psi,
\label{1.6a}
\\[2mm]
&\Phi^p\Psi^{p(p-1)/2}=p^{\ptl^2_t/2}\Phi.
\label{1.6b}
\end{align}
\label{1.6}
\end{subequations}

\noindent
We change the arguments of the fields,
\begin{equation}
\psi(t)=\Psi(t\sqrt{2\ln p}),\qquad\vfi(t)=\Phi(t\sqrt{2\ln p}).
\label{1.7}
\end{equation}
In the class of measurable functions $(\psi,\vfi)$ satisfying growth
condition~\eqref{5.2} (see below) for $\gamma=1$ or $\gamma=2$,
system~\eqref{1.6} becomes the classical system of nonlinear integral
equations
\begin{subequations}
\begin{align}
&\psi^{p^2}(t)=\sqrt{\frac2{\pi}}\int_{-\infty}^\infty
e^{-2(t-\tau)^2}\psi(\tau)\,d\tau,\quad t\in\R,
\label{1.8a}
\\[2mm]
&\vfi^p(t)\psi^{p(p-1)/2}(t)=\frac1{\sqrt\pi}\int_{-\infty}^\infty
e^{-(t-\tau)^2}\vfi(\tau)\,d\tau,\quad t\in\R.
\label{1.8b}
\end{align}
\label{1.8}
\end{subequations}

\noindent
Equation~\eqref{1.4} (Eq.~\eqref{1.8b} for $\psi=1$) for an open string also
assumes a similar form,
\begin{equation}
\vfi^p(t)=\frac1{\sqrt\pi}\int_{-\infty}^\infty
e^{-(t-\tau)^2}\vfi(\tau)\,d\tau,\quad t\in\R.
\label{1.9}
\end{equation}

In accordance with vacuum solutions~\eqref{1.5}, we set the boundary
conditions
\begin{equation}
\lim_{t\to-\infty}\psi(t)=\lim_{t\to\infty}\psi(t)=1
\label{1.10}
\end{equation}
for Eq.~\eqref{1.8a} and
\begin{equation}
\lim_{t\to\infty}\vfi(t)=1,\qquad
\lim_{t\to-\infty}\vfi(t)=\begin{cases}-1&\text{for odd }p,\\[1mm]
\phantom{-}0&\text{for even }p\end{cases}
\label{1.11}
\end{equation}
for Eqs.~\eqref{1.8b} and~\eqref{1.9}.

System~\eqref{1.3} is nonlinear and involves pseudodifferential terms with
the symbols $p^{-\xi^2/4}$ and $p^{-\xi^2/2}$, $\xi^2=t^2-x_1^2-\cdots-
x_{d-1}^2$. It is therefore natural to consider them in some algebras of
generalized functions in $\mathfrak D'(\R^d)$ whose Fourier transforms are
analytic functionals in the space $Z'$~\cite{3},~\cite{4}. Only real
solutions of system~\eqref{1.3} are physically interesting, and we consider
precisely these solutions in what follows.

If $(\Psi(t,x)$, $\Phi(t,x))$ is a solution of system~\eqref{1.3}, then all
its shifts and all its reflections are also solutions of it. If $p$ is an odd
number of the form $p=4n+1$, then $(-\Psi(t,x),-\Phi(t,x))$ is also a
solution of the system. Therefore, the solution of system~\eqref{1.3} is not
unique (if it exists).

Many investigations of physicists and mathematicians widely applying computer
techniques are devoted to this new class of equations with an infinite number
of derivatives (see~\cite{1},~\cite{2},~\cite{4}--\cite{14}, and the
references therein). The interaction is nonlocal in string field
theory~\cite{5}, essentially distinguishing it from the classical local field
theory. These equations are extremely interesting not only for $p$-adic
mathematical physics but also for cosmology~\cite{7},~\cite{9},~\cite{13}. In
essence, these problems relate to the classical mathematical analysis because
only an integer $p$ remains here as one of the $p$-adic numbers, which,
moreover, need not be prime.

In this paper, we study the structure of solutions for open, closed, and
open-closed strings in the framework of the suggested model. In
Sec.~\ref{sec2}, we list some well-known mathematical results for the open
string (boundary value problem~\eqref{1.9},~\eqref{1.11}). In
Sec.~\ref{sec3}, we transfer many results for an open string to the closed
string (boundary value problem~\eqref{1.8a},~\eqref{1.10}) practically
unchanged. For odd $p$, we use the method of successive approximations for
even solutions with two zeros. For even $p$, we prove (Theorem~\ref{th1};
also see~\cite{8}) that there are no continuous even nontrivial solutions
nondecreasing for $t>0$. Therefore, continuous solutions in this case must
either have at least four zeros or be discontinuous with jumps of the first
kind. In Sec.~\ref{sec4}, we prove that the method of successive
approximations converges for an open-closed string (boundary value
problem~\eqref{1.8},~\eqref{1.10},~\eqref{1.11}) with $p$ of the form
$p=4n+1$ (Theorem~\ref{th2}) in the case of boundary value
problem~\eqref{1.8b},~\eqref{1.11} for an odd solution having one zero under
the condition that an even solution of boundary value
problem~\eqref{1.8a},~\eqref{1.10} is known. We describe the solution
structure. In Sec.~\ref{sec5}, we give some necessary properties of the
integral operator $K_\gamma=e^{1/(4\gamma)\ptl^2_t}$.

\section{Open string}
\label{sec2}

The dynamics of an open string are determined by integral equation~\eqref{1.9}
and boundary conditions~\eqref{1.11}. The following propositions hold for
solutions $\vfi$ of Eq.~\eqref{1.9}.

\begin{proposition}
\label{prop1}
If the solution $\vfi(t)$ is bounded, then the function $\vfi(t)=0,\pm1$ for
odd $p$ and the function $\vfi(t)=0,1$ for even $p$ are solutions. If the
solution $\vfi(t)$ does not reduce to a constant, then it is piecewise
analytic {\rm(}continuous for odd $p${\rm)} with the estimate
\begin{equation}
|\vfi(t)|<1,\quad t\in\R.
\label{2.1}
\end{equation}
\end{proposition}

\begin{proposition}
\label{prop2}
If $\vfi(t)\to a$ as $t\to\infty$, $|a|<\infty$, then $a=0$ or $a=\pm1$
for odd $p$, and $a=0$ or $a=1$ for even $p$. In this case,
$(\vfi^p)'(t)\to0$ as $t\to\infty$, and if $a\ne0$, then $\vfi'(t)\to0$ as
$t\to\infty$.
\end{proposition}

\begin{proposition}
\label{prop3}
Integral equation~\eqref{1.9} is equivalent to the boundary value problem
\begin{align}
&u_x=\frac14u_{tt},\quad 0<x\le1,\quad t\in\R,
\label{2.2}
\\[2mm]
&u(0,t)=\vfi(t),\qquad u(1,t)=\vfi^p(t),\quad t\in\R,
\label{2.3}
\end{align}
for the heat conduction equation~{\rm\cite{11}.}
\end{proposition}

We point out that the variables $x$ and $t$ in Eq.~\eqref{2.2} have been
interchanged compared with the classical heat conduction equation.

By a solution of boundary value problem~\eqref{2.2},~\eqref{2.3}, we mean
any measurable function $u(x,t)$ satisfying growth condition~\eqref{5.2}
with respect to $t$ for $\gamma=1$. The function $u(x,t)$ is called an
{\sl interpolating} function between the solution $\vfi$ and its $p$th power
$\vfi^p$. We note that the interpolating function is given by the Poisson
formula for Eq.~\eqref{2.2},
\begin{equation}
u(x,t)=\frac1{\sqrt{\pi x}}\int_{-\infty}^\infty\vfi(\tau)
e^{-(t-\tau)^2/x}\,d\tau,\quad0<x\le1.
\label{2.4}
\end{equation}
The following proposition holds for the zeros of the interpolating function
$u(x,t)$.

\begin{proposition}[{\rm(Theorem on the branching of zeros for the function
$u(1,t)=\vfi^p(t)$)~\cite{11},~\cite{14}}]
\label{prop4}
Let the function $u(x,t)$ have a zero of even multiplicity $2n$ at the point
$t=0$. Then the equation
\begin{equation}
u(1-\vep,t)=0\quad\text{as }\vep\to+0
\label{2.5}
\end{equation}
has exactly $2n$ distinct simple real zeros,
\begin{equation}
t_k^\pm(\vep)=\pm\lambda_k\sqrt\vep+O(\vep),\quad k=1,2,\dots,n,
\label{2.6}
\end{equation}
where $\lambda_k$, $k=1,2,\dots,n$ are positive roots of the Hermite
polynomial, $H_{2n}(\lambda)=0$.
\end{proposition}

For example, if $n=2$, then we have $\lambda^4-12\lambda^2+12=0$, and hence
$$
\lambda_1=\sqrt{6-2\sqrt6}\approx1.049,\qquad
\lambda_2=\sqrt{6+2\sqrt6}\approx3.301.
$$

\begin{proposition}
\label{prop5}
If $\vfi\in\mathfrak L_2^1$, then the solution $\vfi$ expands in a series in
the Hermite polynomials~{\rm\cite{11}},
\begin{equation}
\vfi(t)=\sum_{n=0}^\infty a_n\frac{H_n(t)}{2^nn!},\qquad a_n=(\vfi,H_n)_1,
\label{2.7}
\end{equation}
and the function $\vfi^p(t)$ expands in a Taylor series,
\begin{equation}
\vfi^p(t)=\sum_{n=0}^\infty a_n\frac{t^n}{n!},
\label{2.8}
\end{equation}
which converges uniformly on every compact set in $\R$. And if
$\vfi\in\mathfrak L_2^{1/2}$, then the relations
\begin{equation}
(\vfi^p,H_n)_1=(\vfi,V_n)_{1/2},\quad n=0,1,\dots,
\label{2.9}
\end{equation}
hold, where $V_n$ are modified Hermite polynomials,
\begin{equation}
V_n(t)=2^{-n/2}H_n\biggl(\frac t{\sqrt{2}}\biggr),\quad n=0,1,\dots\,.
\label{2.10}
\end{equation}
\end{proposition}

The space $\mathfrak L_2^{\alpha}$ is defined in Sec~\ref{sec5}.

\begin{proposition}
\label{prop6}
Let $\vfi(t)$ be a solution of Eq.~\eqref{1.9}, and let $t=0$ be a zero of
multiplicity $\sigma\ge1$ for the function $\vfi^p(t)$. Then the relations
\begin{equation}
\frac{2^n}{\sqrt\pi}\int_{-\infty}^\infty\vfi(\tau)\tau^n
e^{-\tau^2}\,d\tau=\begin{cases}0&\text{for }n=0,1,\dots,\sigma-1,\\[1mm]
a\ne0&\text{for }n=\sigma\end{cases}
\label{2.11}
\end{equation}
hold~{\rm\cite{4}}. The function $\vfi(t)$ has at least $\sigma$ sign
changes~{\rm\cite{11}.}
\end{proposition}

\newcounter{newpro}
\def\theproposition{{\bf\arabic{proposition}\alph{newpro}}}

\addtocounter{newpro}{1}
\begin{proposition}
\label{prop6a}
For odd $p$ and $\sigma$, the solution $\vfi(t)$ changes sign at zero,
and the asymptotic relation
\begin{equation}
\vfi(t)=\biggl(\frac{a}{\sigma!}\biggr)^{1/p}t^{\sigma/p}[1+O(t)]\quad
\text{as }t\to0
\label{2.12}
\end{equation}
holds.
\end{proposition}

\setcounter{proposition}{6}
\addtocounter{newpro}{1}
\begin{proposition}
\label{prop6b}
For even $p$, the multiplicity $\sigma$ is even, and we have $a>0$. If
$\vfi(t)$ changes sign at zero, then
\begin{equation}
\vfi(t)=\biggl(\frac a{\sigma!}\biggr)^{1/p}\sign t|t|^{\sigma/p}[1+O(t)]
\quad\text{as }t\to0.
\label{2.13}
\end{equation}
\end{proposition}

\setcounter{proposition}{6}
\addtocounter{newpro}{1}
\begin{proposition}
\label{prop6c}
If $\vfi(t)$ does not change sign at zero, then $a>0$, the multiplicity
$\sigma$ is even, and
\begin{equation}
\vfi(t)=\biggl(\frac a{\sigma!}\biggr)^{1/p}|t|^{\sigma/p}[1+O(t)]\quad
\text{as }t\to0.
\label{2.14}
\end{equation}
\end{proposition}

The following propositions hold for solutions of boundary value
problem~\eqref{1.9}, \eqref{1.11}.

\setcounter{proposition}{7}
\def\theproposition{\arabic{proposition}}

\begin{proposition}
\label{prop7}
For odd $p$, there is a solution $\vfi(t)$ that is continuous, real-analytic
for $t\,{\ne}\,0$, odd, and increasing. It has one simple zero at $t=0$, and
relations~\eqref{2.11} and~\eqref{2.12} hold for
$\sigma=1${\rm~\cite{4},~\cite{10},~\cite{12}.}
\end{proposition}

\begin{proposition}
\label{prop8}
For even $p$, there are no continuous solutions that are either nonnegative
or have only one change of sign. If there is a continuous solution with two
zeros, then it has exactly two sign changes~{\rm\cite{11}.} Here,
discontinuous solutions with discontinuities of the first kind are possible.
\end{proposition}

\begin{proposition}
\label{prop9}
Let $u(x,t)$ be an interpolating function between a solution $\vfi$ and
its $p$th power $\vfi^p$. Then the conservation law
\begin{equation}
\int_{-\infty}^\infty[\vfi(t)-u(x,t)]\,dt=0=
\int_{-\infty}^\infty[\vfi(t)-\vfi^p(t)]\,dt,\quad x\ge0,
\label{2.15}
\end{equation}
and the inequality
\begin{equation}
\biggr|\int_{-\infty}^a[u(x,t)-\vfi(t)]\,dt\biggl|<
\sqrt{\frac x{\pi}},\quad x>0,\quad a\in\R,
\label{2.16}
\end{equation}
hold.
\end{proposition}

\begin{proposition}
\label{prop10}
For odd $p$, the inclusions
\begin{equation}
1-\vfi^{p-1},1-|u(x,\,\cdot\,)|\in\mathfrak L_1(\R),\quad x\ge0,
\label{2.17}
\end{equation}
hold, the function $\vfi^p(t)$ has finitely many zeros, and all its zeros
have a finite multiplicity. The number of sign changes of $\vfi(t)$ is not
greater than the number of zeros for $\vfi^p(t)$ and not less than the maximum
multiplicity of zeros for this function. If $\vfi^p(t)$ has only three zeros,
then the solution $\vfi(t)$ has three changes of sign~{\rm\cite{11}.}
\end{proposition}

\begin{proposition}
\label{prop11}
For even $p$, the inclusions
\begin{equation}
1-\vfi^{p-1},1-u(x,\,\cdot\,)\in\mathfrak L_1(0,\infty),\quad x\ge0,
\label{2.18}
\end{equation}
hold, the integral
\begin{equation}
\int_{-\infty}^0\vfi(t)[1-\vfi^{p-1}(t)]\,dt
\label{2.19}
\end{equation}
converges, and if the function $\vfi(t)$ has a constant sign for $t<c$, then
the inclusions
\begin{equation}
\vfi,u(x,\,\cdot\,)\in\mathfrak L_1(-\infty,0),\quad x\ge0,
\label{2.20}
\end{equation}
hold.
\end{proposition}

The set of zeros of the function $\vfi^p(t)$ is finite or countable and
bounded above. We let $t_0,t_1,\dots,t_k\to -\infty$ denote these zeros
and $\sigma_0,\sigma_1,\dots$ denote their multiplicities~\cite{11}. By the
Hadamard theorem (Lemma~\ref{lem2} in Sec.~\ref{sec5}), the inequality
$$
\sum_{k=0}^\infty\sigma_k|t_k|^{-2-\vep}<\infty\quad
\text{for an arbitrary }\vep>0
$$
holds.

Here, the following questions arise.

\vspace*{-2mm}
\begin{list}{}{\setlength{\leftmargin}{12mm}%
\setlength{\itemindent}{0pt}}
\item[1.]Does a continuous or discontinuous solution of boundary value
problem~\eqref{1.9},~\eqref{1.11} exist for even~$p$?

\item[2.]Does a change of sign of the solution $\vfi(t)$ at a zero of the
function $\vfi^p(t)$ always occur?

\item[3.]Are the multiplicities of zeros of the function $\vfi^p(t)$ only odd
for odd $p$ and only even of the form $2(2n+1)$ for even $p$?
\end{list}

\section{Closed string}
\label{sec3}

The dynamics of a closed string are determined by integral
equation~\eqref{1.8a} and boundary conditions~\eqref{1.10}.
Equation~\eqref{1.8a} reduces to Eq.~\eqref{1.9} by replacing $p$ with $p^2$
and the operator $K_1$ with $K_2$ (see Sec.~\ref{sec5}). Therefore,
Propositions~\ref{prop1}--\ref{prop6} in Sec.~\ref{sec2} relating to bounded
solutions of Eq.~\eqref{1.9} also hold for Eq.~\eqref{1.8a}.
Propositions~\ref{prop8}--\ref{prop10} in Sec.~\ref{sec2} hold for boundary
value problem~\eqref{1.8a},~\eqref{1.10}, and Propositions~\ref{prop7}
and~\ref{prop11} are replaced with the following propositions.

\setcounter{proposition}{7}
\def\theproposition{\arabic{proposition}$'$}

\begin{proposition}
\label{prop7'}
There are no nonnegative continuous solutions except\linebreak
$\psi(t)=1$. For even
$p$, discontinuous solutions with discontinuities of the first kind are
possible~{\rm\cite{4}.}
\end{proposition}

\setcounter{proposition}{11}
\def\theproposition{\arabic{proposition}$'$}

\begin{proposition}
\label{prop11'}
The function $\psi^{p^2}$ has finitely many zeros for both odd and even $p$,
and  the inclusions
\begin{equation}
1-\psi^{p^2-1},\,1-u(x,\,\cdot\,)\in\mathfrak L_1(\R),\quad x\ge0,
\label{3.1}
\end{equation}
therefore hold.
\end{proposition}

To construct an approximate solution of boundary value
problem~\eqref{1.8a}, \eqref{1.10}, we use the method of successive
approximations elaborated in~\cite{4} and~\cite{10} for an open string in the
case of odd $p$. We seek an even solution $\psi(t)$ with two zeros $\pm t_0$.
Let the successive approximations be given by the recursive formula
\begin{equation}
\psi_{n+1}(t)=[(K_2\psi_n)(t)]^{1/p^2},\quad n=0,1,\dots,\qquad
\psi_0(t)=1-\beta e^{-\alpha t^2}.
\label{3.2}
\end{equation}
Here, the operator $K_2$ is determined by formula~\eqref{5.1} (see below)
for $\gamma=2$. For an even function $\psi$, it becomes
$$
(K_2\psi)(t)=\sqrt{\frac2{\pi}}\int_0^\infty\psi(\tau)
[e^{-2(t-\tau)^2}+e^{-2(t+\tau)^2}]\,d\tau.
$$
We calculate the second approximation. We have
$$
(K_2\psi_0)(t)=\sqrt{\frac2{\pi}}\int_{-\infty}^\infty [1-\beta
e^{-\alpha\tau^2-2(t-\tau)^2}]\,d\tau=
1-\beta\sqrt{\frac2{\alpha+2}}e^{-2\alpha t^2/(\alpha+2)},
$$
whence we derive the formula
$$\psi_1(t)=\biggl[1-\beta\sqrt{\frac2{\alpha+2}}
e^{-2\alpha t^2/(\alpha+2)}\biggr]^{1/p^2}
$$
by~\eqref{3.2}. Finally,
\begin{equation}
\psi_2(t)=\biggl\{\sqrt{\frac2{\pi}}
\int_0^\infty\biggl[1-\beta\sqrt{\frac2{\alpha+2}}
e^{-2\alpha\tau^2/(\alpha+2)}\biggr]^{1/p^2}\!\!\!
[e^{-2(t-\tau)^2}+e^{-2(t+\tau)^2}]\,d\tau\biggr\}^{1/p^2}\!\!\!.
\label{3.3}
\end{equation}
The function $\psi_2(t)$ constructed using formula~\eqref{3.3}, as well as
the function $\psi_1(t)$, is a good approximation to the solution of boundary
value problem~\eqref{1.8a}, \eqref{1.10}. To find the zeros $\pm t_0$ of
$\psi_2^{p^2}(t)$, we must solve the equation
\begin{equation}
\int_0^\infty\biggl[1-\beta\sqrt{\frac2{\alpha+2}}
e^{-2\alpha\tau^2/(\alpha+2)}\biggr]^{1/p^2}
[e^{-2(t-\tau)^2}+e^{-2(t+\tau)^2}]\,d\tau=0.
\label{3.4}
\end{equation}

\begin{examples}
Let $p=3$ and $\alpha=0,1$. Then

\vspace*{-2mm}
\begin{list}{}{\setlength{\leftmargin}{12mm}%
\setlength{\itemindent}{0pt}}
\item[1.]$\psi_0(0)=-0.500$, $t^0_0=2.01$, $\psi_1(0)=-0.917$, and
$t^1_0=2.00$ for $\beta=1.5$,

\item[2.]$\psi_0(0)=-0.800$, $t^0_0=2.42$, $\psi_1(0)=0.971$, and
$t^1_0=2.43$ for $\beta=1.8$,

\item[3.]$\psi_0(0)=-0.900$, $t^0_0=2.53$, $\psi_1(0)=-0.982$, and
$t^1_0=2.55$ for $\beta=1.9$, and

\item[4.]$\psi_0(0)=-1$, $t^0_0=2.63$, $\psi_1(0)=-0.995$, and
$t^1_0=2.65$ for $\beta=2$.
\end{list}
\end{examples}

\begin{figure}[t]
\centering
\includegraphics[height=66mm]{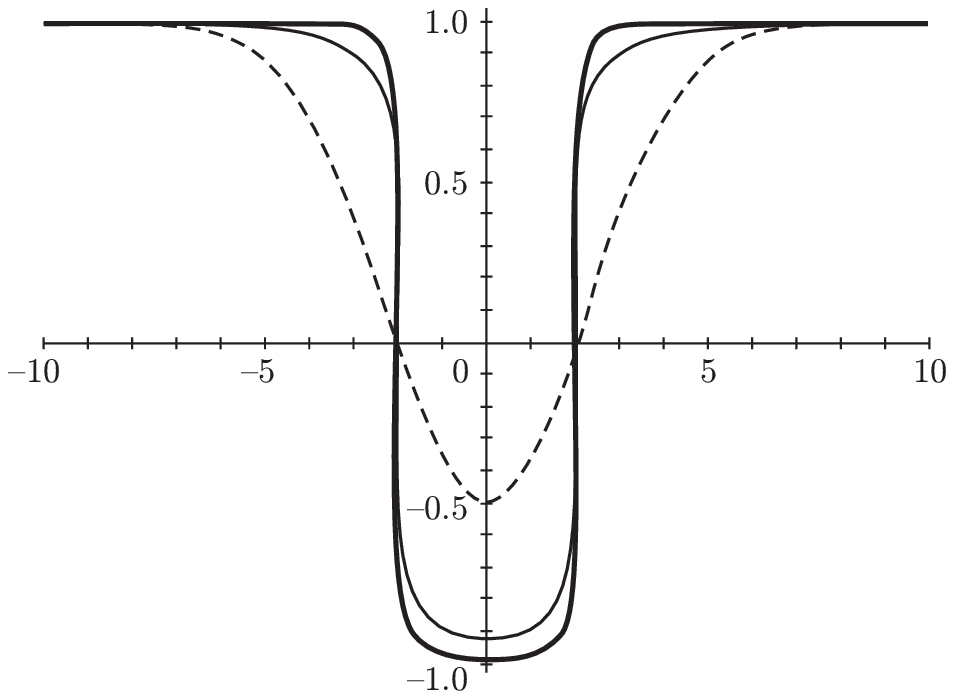}
\caption{}
\label{fig1}
\end{figure}

\begin{figure}[t]
\centering
\includegraphics[height=66mm]{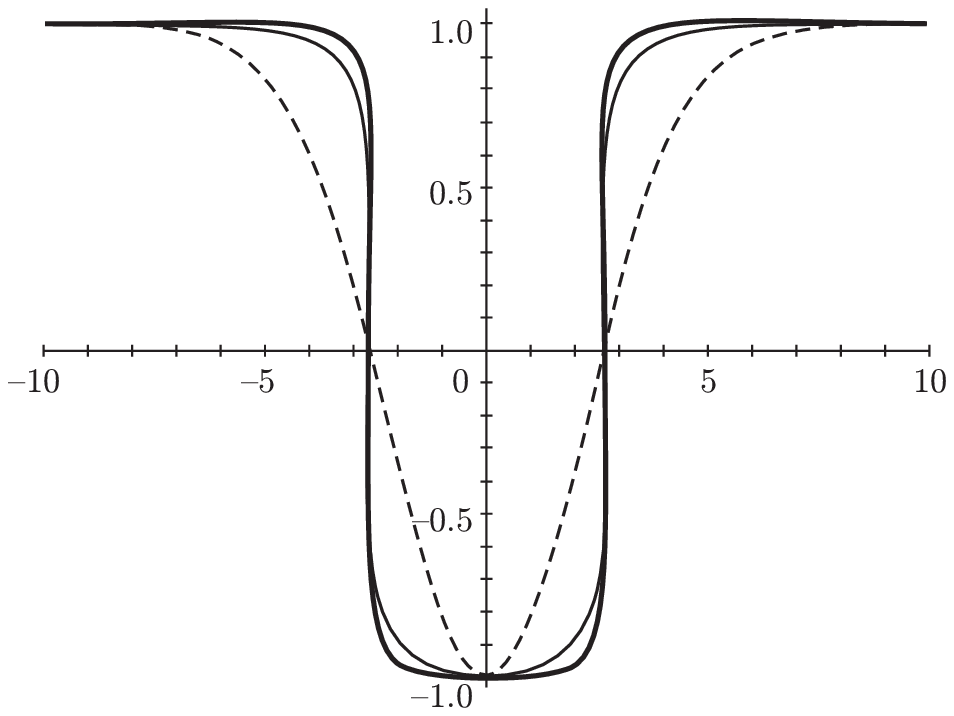}
\caption{}
\label{fig2}
\end{figure}

Figure~\ref{fig1} ($\beta=1.5$, $t^n_0\approx 2.00$) and Fig.~\ref{fig2}
($\beta=2$, $t^n_0\approx 2.65$) demonstrate the iterations $\psi^9_{n+1}=
(K_2\psi_n)^{1/9}$. Here, $\psi_0$ is represented by the dashed line;
$\psi_1$, by the thin continuous line; and $\psi_2$, by the heavy line.

As is shown by numerical calculations (see the examples and the figures), for
every $\beta$ in the interval $1.5\le\beta\le2$, the zeroth approximation
$\psi_0=1-e^{-0.1t^2}$ determines a $\beta$-dependent solution to which the
iterations rapidly converge (and the zeros $t^n_0$, $n=0,1,2$, are therefore
indistinguishable in Figs.~\ref{fig1} and~\ref{fig2}). We hence have a
one-parameter family of approximate solutions depending on the parameter
$\beta$ in the initial function $\psi_0$. But the convergence proof for
successive approximations demonstrated in~\cite{4},~\cite{10}, and~\cite{12}
for an open string does not apply in this case. Nevertheless, the following
proposition holds.

\def\theproposition{\arabic{proposition}}

\begin{proposition}
\label{prop12}
Let $p$ be odd. If there is a continuous even solution $\psi(t)$ with the two
zeros $t=\pm t_0$ for the boundary value problem~\eqref{1.8a},~\eqref{1.10},
then these zeros are simple, and the relations
\begin{equation}
\psi(t)=[a(t\pm t_0)]^{p^{-2}}[1+O(|t\pm t_0|)],\quad t\to\pm t_0,
\label{3.5}
\end{equation}
hold, where
\begin{equation}
a=\sqrt{\frac2{\pi}}\int_0^\infty\psi(\tau)[e^{-2(t_0-\tau)^2}+
e^{-2(t_0+\tau)^2}]\,d\tau.
\label{3.6}
\end{equation}
\end{proposition}

\begin{proof}
We show that the zeros $\pm t_0$ are simple. Indeed, if the zero $t_0$ is
multiple, then the first derivative of $\psi^{p^2}(t)$ must vanish at the
point $t_0$. But $\psi(t)$ and consequently $\psi^{p^2}(t)$ change sign at
$t_0$. Therefore, the second derivative of $\psi^{p^2}(t)$ also
vanishes at this point. This means that the multiplicity of the zero $t_0$ of
this function is not less than three, and  the number of its sign changes must
therefore be not less than three (see Proposition~\ref{prop6}), which
contradicts our assumption.
\end{proof}

The following negative result, first stated in~\cite{8}, holds for boundary
value problem~\eqref{1.8a},~\eqref{1.10} with even values of $p$.

\begin{theorem}
\label{th1}
There are no continuous even solutions $\psi$ nondecreasing for $t>0$ except
$\psi\equiv1$.
\end{theorem}

\begin{proof}
Let $\psi(t)\not\equiv1$ be a continuous even solution of boundary value
problem~\eqref{1.8a},~\eqref{1.10}, and let $\psi(t)$ be nondecreasing for
$t>0$. Then $\psi(0)<0$ because the function $\psi(t)$ would otherwise be
nonnegative for all $t\in\R$ by our assumption, which would contradict
Proposition~\ref{prop7'}. Therefore, there is a point $t=t_0$ such that
$\psi(t)<0$ for all $t$, $0\le t<t_0$, and $\psi(t_0)=0$. On the other hand,
the function $\psi'(t)$, $t\ge0$, is nonnegative and continuous everywhere
except at $t_0$, where it has an integrable singularity (see
Proposition~7), Furthermore, $\psi'(\infty)=0$ (see Proposition~\ref{prop2}).
According to Lemma~\ref{lem4} (see Sec.~\ref{sec5}), the relation
$$
p^2\psi^{p^2-1}(t)\psi'(t)=\sqrt{\frac2\pi}\int_0^\infty\psi'(\tau)
[e^{-2(t-\tau)^2}-e^{-2(t+\tau)^2}]\,d\tau\ge0
$$
holds, which leads to a contradiction for $0<t<t_0$ because $p^2-1$ is an odd
number, and therefore $\psi^{p^2-1}(t)<0$ for $0<t<t_0$.
\end{proof}

The meaning of the above proposition is that nontrivial even solutions of
boundary value problem~\eqref{1.8a},~\eqref{1.10} must either have an even
number of zeros not less than four (see Proposition~\ref{prop10}) or be
discontinuous.

\section{Open-closed string}
\label{sec4}

The dynamics of an open-closed string are determined by system~\eqref{1.8}
and boundary conditions~\eqref{1.10} and~\eqref{1.11}. Boundary value
problem~\eqref{1.8a},~\eqref{1.10} describes a closed string, and the
solution of this problem was discussed in Sec.~\ref{sec3}. Let $\psi_0(t)$ be
a known even solution of~\eqref{1.8a},~\eqref{1.10} such that the
multiplicities $\sigma_k$, $k=1,2,\dots,m$, of zeros of $\psi_0^{p^2}(t)$
satisfy the inequality
\begin{equation}
\sigma_k<\frac{2p^2}{p-1},\quad k=1,2,\dots,m.
\label{4.1}
\end{equation}
Substituting this solution in Eq.~\eqref{1.8b}, we obtain the equation
\begin{equation}
\vfi^p(t)\psi_0^{p(p-1)/2}(t)=(K_1\vfi)(t),
\label{4.2}
\end{equation}
where the operator $K_1$ is given by formula~\eqref{5.1} for $\gamma=1$. To
solve Eqs.~\eqref{4.2}, we introduce a new unknown function,
\begin{equation}
\chi(t)=\vfi(t)\psi_0^{(p-1)/2}(t),\qquad\vfi(t)=\chi(t)\psi_0^{-(p-1)/2}(t).
\label{4.3}
\end{equation}
Substituting~\eqref{4.3} in Eq.~\eqref{4.2}, we obtain the integral equation
\begin{equation}
\chi^p(t)=(K_1 v\chi)(t),
\label{4.4}
\end{equation}
where we introduce the notation
\begin{equation}
v(t)=\psi_0^{-(p-1)/2}(t).
\label{4.5}
\end{equation}
The function $v$ has the following properties: $|v(t)|>1$, $t\in\R$; it is
even; it is (real-)analytic everywhere except at finitely many zeros of
$\psi_0(t)$, where it has an integrable singularity by condition~\eqref{4.1}
(see Propositions~7a and~7c with $p$ replaced with $p^2$)
and satisfies boundary conditions~\eqref{1.10}; and by~\eqref{3.1} (because
$p^2-1>(p-1)/2$), the inclusions
\begin{equation}
v-1,\,|v|-1\in\mathfrak L_1(\R)
\label{4.6}
\end{equation}
hold. We prove the estimate
\begin{equation}
(K_1|v|)(t)<N,\quad t\in\R,\qquad
N=1+\frac2{\sqrt\pi}\int_0^\infty\bigl[|v(\tau)|-1\bigr]\,d\tau.
\label{4.7}
\end{equation}
Indeed, by~\eqref{4.6}, we have
$$
(K_1|v|)(t)<\frac1{\sqrt\pi}\int_{-\infty}^\infty
\bigl[|v(\tau)|-1\bigr]e^{-(t-\tau)^2}\,d\tau+1.
$$
Estimate~\eqref{4.7} implies the inequality
\begin{equation}
|\chi(t)|<N^{1/(p-1)},\quad t\in\R.
\label{4.8}
\end{equation}
Indeed, Eq.~\eqref{4.4} implies
$$
|\chi^p(t)|=|\chi(t)|^p=|K(v\chi)(t)|<
\max_{t\in\R}|\chi(t)|(K|v|)(t)<N\max_{t\in\R}|\chi(t)|,
$$
whence~\eqref{4.8} precisely follows. It follows from~\eqref{4.3}
and~\eqref{4.8} that the solution satisfies the estimate
\begin{equation}
|\vfi(t)|<N^{1/(p-1)}|\psi_0^{-(p-1)/2}(t)|,\quad t\in\R.
\label{4.9}
\end{equation}

We now assume that $p$ is an odd number of the form $p=4n+1$. In this case,
we have $v(t)>1$, $t\in\R$, by~\eqref{4.5}. As in the case of an open
string~\cite{4},~\cite{10}, we seek an odd solution $\chi(t)$ of boundary
value problem~\eqref{4.4},~\eqref{1.11} by the method of successive
approximations,
\begin{equation}
\chi_{n+1}(t)=\bigl[(K_1v\chi_n)(t)\bigr]^{1/p},\quad n=0,1,\dots,\qquad
\chi_0(t)=\sign t,\quad t\in\R.
\label{4.10}
\end{equation}
The approximations $\chi_n(t)$, $n=1,2,\dots$, are odd, continuous, and
positive functions. They increase for $t>0$, vanish at the point $t=0$, and
tend to unity as $t\to\infty$. Every entire function $\chi_n^p(t)$,
$n=1,2,\dots$, has a simple zero at $t=0$. The simplicity of the zero follows
from~\eqref{4.10} in view of the relations
\begin{equation}
\frac d{dt}\chi_n^p(0)=\frac d{dt}(K_1 v\chi_{n-1})(0)=
\frac4{\sqrt\pi}\int_0^\infty v(\tau)\chi_{n-1}(\tau)
\tau e^{-\tau^2}\,d\tau>0.
\label{4.11}
\end{equation}

We prove that there are positive numbers $\eta$ and $\theta$ such that
\begin{equation}
\eta\chi_1(t)\le\chi_2(t)\le\theta\chi_1(t),\quad t\ge0.
\label{4.12}
\end{equation}
We introduce the function $f(t)=\chi_2^p(t)\chi_1^{-p}(t)$. It is continuous
and positive for $t>0$ and tends to unity as $t\to\infty$. By~\eqref{4.11},
its limit as $t\to0$ exists and (according to L'Hospital's rule) is finite.
Consequently, there are some numbers $a$ and $b$, $0<a<b$, such that
$$
a\le f(t)\le b,\quad t\ge0,
$$
whence inequality~\eqref{4.12} for $\eta=a^{1/p}$ and $\sigma=b^{1/p}$
precisely follows.

Multiplying~\eqref{4.12} by $v$, applying the operator $K_1$, and recalling
that the kernel of $K_1$ is nonnegative, we obtain the inequalities
$$
\eta(K_1 v\chi_1)(t)\le(K_1 v\chi_2)(t)\le\theta(K_1 v\chi_1)(t),
$$
whence we use~\eqref{4.10} to derive the inequality
$$
\eta\chi_2^p(t)\le\chi_3^p(t)\le\theta\chi_2^p(t),
$$
consequently
$$
\eta^{1/p}\chi_2(t)\le\chi_3(t)\le\theta^{1/p}\chi_2(t),
$$
and so on. As a result, we obtain the inequalities
$$
\eta^{p^{-n+1}}\chi_n(t)\le\chi_{n+1}(t)\le
\theta^{p^{-n+1}}\chi_n(t),\quad n=2,3,\dots,\quad t\ge0.
$$
Arguing as in~\cite{4}, we now conclude that the sequence of iterations
$\chi_n(t)$, $n=0,1,\dots$, converges uniformly on $\R$ to the solution
$\chi(t)$ of boundary value problem~\eqref{4.3},~\eqref{1.11} for $p$ of
the form $p=4n+1$. We have thus proved the following theorem.

\begin{figure}[t]
\centering
\includegraphics[height=77mm]{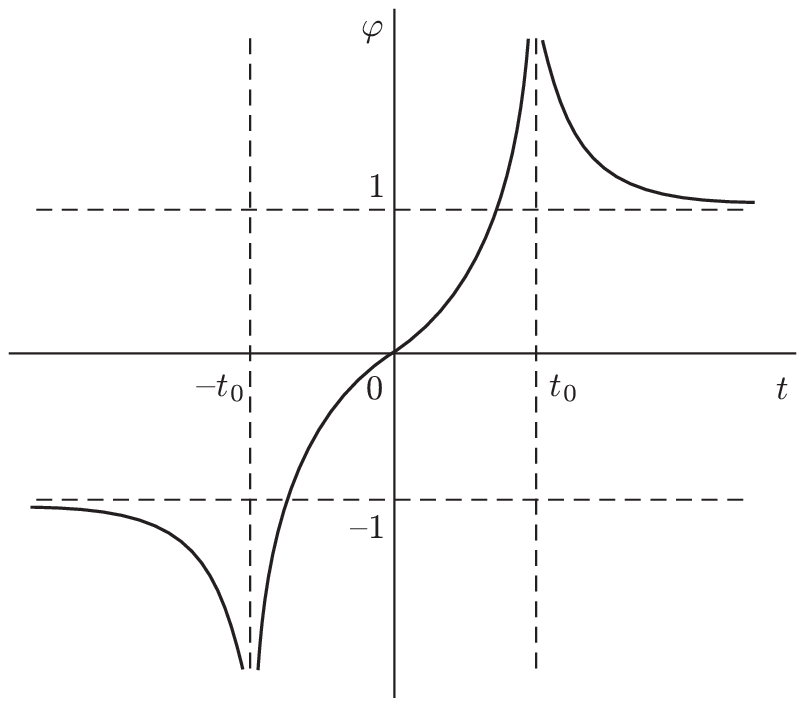}
\caption{}
\label{fig3}
\end{figure}

\begin{theorem}
\label{th2}
Let $p\equiv1\pmod4$, let $\psi_0$ be an even solution of boundary value
problem~\eqref{1.8a},~\eqref{1.10}, let the multiplicities of the zeros be
$\sigma_k$, $k=1,2,\dots,m$, and let the functions $\psi_0^{p^2}(t)$ satisfy
condition~\eqref{4.1}. Then the solution of boundary value
problem~\eqref{1.8},~\eqref{1.10},~\eqref{1.11} for the open-closed string
exists and is given by
\begin{equation}
(\psi=\psi_0,\ \vfi=\chi\psi_0^{-(p-1)/2})
\label{4.13}
\end{equation}
with the estimate
\begin{equation}
|\vfi(t)|\le C\psi_0^{-(p-1)/2}(t),\quad t\in\R,\quad
1-|\vfi|\in\mathfrak L_1(\R),
\label{4.14}
\end{equation}
where $\chi$ is an odd solution of boundary value
problem~\eqref{4.3},~\eqref{1.11} with one zero and the constant $C$ depends
only on $\psi_0$ {\rm(}see~Fig.~\ref{fig3}{\rm).}
\end{theorem}

\begin{remark}
In Theorem~\ref{th2}, each of the even solutions of boundary value
problem~\eqref{1.8a},~\eqref{1.10} that were described in Sec.~\ref{sec3} can
be taken as the solution $\psi_0$.
\end{remark}

In the case of a number $p$ of the form $p=4n+3$, the proof that the
presented method converges does not apply, but successive
approximations~\eqref{4.9} seem to converge.

\section{Properties of the integral operator $K_\gamma$}
\label{sec5}

The integral operator $K_\gamma$ is defined by the formula
\begin{equation}
(K_\gamma f)(t)=\sqrt{\frac{\gamma}{\pi}}\int_{-\infty}^\infty
f(\tau)e^{-\gamma(t-\tau)^2}\,d\tau,\quad\gamma>0,
\label{5.1}
\end{equation}
on the class $\mathfrak B_\gamma$ of locally integrable functions $f$
satisfying the growth condition
\begin{equation}
|f(t)|=O(e^{\vep t^2}),\quad|t|\to\infty,\quad0\le\vep<\gamma.
\label{5.2}
\end{equation}

We need the scale of weighted separable Hilbert spaces
$\mathfrak L_2^\alpha$, $\alpha>0$, consisting of measurable functions square
summable on $\R$ with respect to the measure
$$
d\mu_\alpha(t)=\sqrt{\frac{\alpha}{\pi}}e^{-\alpha t^2}\,dt,\qquad
\int_{-\infty}^\infty\,d\mu_\alpha(t)=1,
$$
with the inner product and norm
$$
(f,g)_\alpha=\int_{-\infty}^\infty f(t)\bar g(t)\,d\mu_\alpha(t),\qquad
\|f\|_\alpha=\sqrt{(f,f)_\alpha},\quad f,g\in\mathfrak L_2^\alpha.
$$

\begin{lemma}
\label{lem1}
The operator $K_\gamma$ maps $\mathfrak L_2^\alpha$ into
$\mathfrak L_2^\beta$ for
$$
0<\alpha<2\gamma,\qquad\beta>\frac{2\alpha\gamma}{2\gamma-\alpha}
$$
and is bounded,
\begin{equation}
\|Kf\|_\beta\le\biggl(\frac{\beta\gamma^2}{2\alpha\beta\gamma-
2\alpha^2\gamma-\beta\alpha^2}\biggr)^{-1/4}\|f\|_\alpha,\quad
f\in L_2^\alpha.
\label{5.3}
\end{equation}
\end{lemma}

\begin{lemma}
\label{lem2}
The operator $K_\gamma$ maps $\mathfrak L_2^\alpha$, $0<\alpha<2\gamma$, into
an entire function $(Kf)(z)$, $z=t+iy$, with a growth order not higher than
the second with estimate
\begin{equation}
|(Kf)(z)|\le\|f\|_\alpha\sqrt\gamma(2\gamma-\alpha)^{-1/4}
\exp\biggl(y^2-\gamma t^2+\frac2{2\gamma-\alpha}t^2\biggr).
\label{5.4}
\end{equation}
\end{lemma}

\begin{lemma}
\label{lem3}
The operator $K_\gamma$ maps a bounded function $f(t)$, $|f(t)|<C$, into a
bounded function $(K_\gamma f)(t)$, $|(K_\gamma f)(t)|<C$, and if $f(t)\to a$
as $t\to\infty$, then $(K_\gamma f)(t)\to a$ as $t\to\infty$.
\end{lemma}

\begin{lemma}
\label{lem4}
If $f,f'\in\mathfrak B_\gamma$ and if $f(t)$ is an odd or even continuous
function for $t\ge0$ continuously differentiable everywhere except at
finitely many isolated points in whose neighborhood $f'(t)$ is integrable,
then the real-analytic function $(K_\gamma f)(t)$ is respectively odd or even
and increases for $t\ge0$.
\end{lemma}

\begin{demo}
For $\gamma=1$, Lemmas~\ref{lem1} and~\ref{lem2} were proved in~\cite{11}
and Lemma~\ref{lem3} in~\cite{4}. They are proved similarly for $\gamma\ne1$.
We prove that $(K_\gamma f)(t)$ is an increasing function for $t\ge0$. By
assumption, $f(t)$ contains no singular part, the derivative $f'(t)$
therefore satisfies the condition $f'(t)\ge0$ almost everywhere, and the
formula for integrating by parts holds for a product with smooth functions.
Using the rule for differentiating a convolution~\cite{3}, we obtain the
inequality
$$
(K_\gamma f)'(t)=\sqrt{\frac{\gamma}{\pi}}\int_0^\infty
f'(\tau)[e^{-\gamma(t-\tau)^2}+e^{-\gamma(t+\tau)^2}]\,d\tau\ge0
$$
for odd functions $f(t)$, $t\ge0$. A similar argument can also be used for
even functions $f$.
\end{demo}

\subsection*{Acknowledgments}
The author expresses his heartfelt gratitude to I.~Ya.~Aref'eva for the useful
discussion and for the numerical calculations.

\smallskip

This work was supported in part by the Program for Supporting Leading
Scientific Schools (Grant No.~NSh-1542.2003.1).


\begin{thebibliography}{99}

\bibitem{1}
L.~Brekke and P.~G.~O.Freund,
Phys. Rep.,
{\bf 233} (1993),
1.

\bibitem{2}
N.~Moeller and M.~Schnabl,
JHEP, {\bf 0401} (2004),
011.

\bibitem{3}
I.~M.~Gelfand and G.~E.~Shilov,
{\it Generalized Functions and Operations on Them},
vol.~2, Spaces of Fundamental and Generalized Functions,
Moscow,
Fizmatlit,
1958
(in Russian);
{\rm English transl.:} Generalized Functions,
vol.~2,
Spaces of Fundamental and Generalized Functions,
New York,
Acad. Press,
1968.

\bibitem{4}
V.~S.~Vladimirov and Ya.~I.~Volovich,
{\it Theor. Math. Phys.},
{\bf 138}
(2004),
297;
{\tt math-ph/0306018},
2003.

\bibitem{5}
M.~B.~Green, J.~H.~Schwarz, and~E.~Witten,
{\it Superstring Theory},
vols.~1,~2,
Cambridge,
Cambridge Univ.~Press,
1987, 1988.

\bibitem{6}
L.~Brekke, P.~G.~O.~Freund, M.~Olson, and E.~Witten,
{\it Nucl. Phys.~B},
{\bf 302} (1988),
365;
V.~S.~Vladimirov, I.~V.~Volovich, and E.~I.~Zelenov,
{\it $p$-Adic Analysis and Mathematical Physics},
Moscow,
Nauka,
1994
(in Russian);
English transl.,
Singapore,
World Scientific,
1994;
A.~Sen
{\it JHEP},
{\bf 0204} (2002),
048;
{\tt hep-th/0203211},
2002;
D.~Ghoshal and A.~Sen,
{\it Nucl. Phys.~B},
{\bf 584} (2000),
300;
I.~V.~Volovich,
{\it Class.~Q.~Grav.},
{\bf 4} (1987),
L83;
J.~A.~Minahan
{\it JHEP},
{\bf 0103} (2001),
028;
N.~Barnaby,
{\it JHEP},
{\bf 0407} (2004),
025;
{\tt hep-th/0406120},
2004;
E.~Coletti, I.~Sigalov, and W.~Taylor,
{\it JHEP},
{\bf 0508} (2005),
104;
{\tt hep-th/0505031},
2005.

\bibitem{7}
P.~H.~Frampton and Y.~Okada,
{\it Phys. Rev.~D},
{\bf 37} (1988),
3077.

\bibitem{8}
N.~Moeller and B.~Zwiebach,
{\it JHEP},
{\bf 0210} (2002),
034;
{\tt hep-th/0207107},
2002.

\bibitem{9}
I.~Ya.~Aref'eva, L.~V.~Joukovskaja, and A.~S.~Koshelev,
{\it JHEP},
{\bf 0309}
(2003),
012;
{\tt hep-th/0301137},
2003.

\bibitem{10}
Ya.~I.~Volovich,
{\it J.~Phys.~A},
{\bf 36} (2003),
8685;
{\tt math-ph/0301028},
2003.

\bibitem{11}
V.~S.~Vladimirov,
{\it Izv. Math.},
{\bf 69} (2005),
487;
{\tt math-ph/0507018},
2005.

\bibitem{12}
L.~V.~Joukovskaja,
{\it Theor. Math. Phys.},
{\bf 146} (2006),
335.

\bibitem{13}
G.~Calcagni,
{\it JHEP},
{\bf 0605} (2006),
012;
{\tt hep-th/0512259},
2005.

\bibitem{14}
V.~S.~Vladimirov,
{\it Russ. Math. Surveys},
{\bf 60}
(2005),
1077.

\end{thebibliography}
\end{document}